\documentclass[aps,prd,reprint,onecolumn,notitlepage,superscriptaddress,nofootinbib,floatfix]{revtex4-2}
\usepackage{amsmath,amssymb,bm}
\usepackage{slashed}
\usepackage{graphicx}
\usepackage{booktabs}
\usepackage{array}
\usepackage{dcolumn}
\usepackage{xcolor}
\usepackage{hyperref}
\hypersetup{
  colorlinks=true,
  citecolor=blue,
  linkcolor=blue,
  urlcolor=blue
}

\newcommand{\Lb}{\Lambda_b}
\newcommand{\Pbar}{\bar P}

\newcommand{\Tr}{\operatorname{Tr}}

\begin{document}

\title{Two-variable longitudinal distribution of the $\Lambda_b$ in a three-quark light-front model}

\author{Shuai Xu}
\email{sxu\_iopp@foxmail.com}
\affiliation{School of Physics and Telecommunications Engineering, Zhoukou Normal University, Zhoukou 466001, People's Republic of China}
\author{Xiao-Nan Li}
\affiliation{School of Electrical and Information Engineering, Tongling University, Tongling 244061, People's Republic of China}
\begin{abstract}
We determine a normalized two-variable longitudinal distribution of the $\Lambda_b=(ud)b$ baryon from a covariant three-quark light-front wave function. The scalar $ud$ channel is selected with the good-component structure $C\gamma_5\gamma^+$ and the two independent transverse momenta are integrated without introducing an elementary diquark. The resulting model-scale valence distribution $\Psi_2^{\rm LF}(x_u,x_d)$ retains both the heavy-light and internal-light momentum scales of the parent wave function. It is symmetric under $x_u\leftrightarrow x_d$, is concentrated where both light-quark fractions are small and separates the total light-pair momentum from its internal sharing. For the central parameter set we find $\langle x_u\rangle=\langle x_d\rangle=0.130$, $\langle x_b\rangle=0.739$, $\langle\zeta^2\rangle=0.153$, and $\langle\zeta^4\rangle=0.055$, where $\zeta=(x_u-x_d)/(x_u+x_d)$. Parameter variations give $\langle x_u\rangle=0.130^{+0.007}_{-0.007}$, $\langle x_b\rangle=0.739^{+0.014}_{-0.014}$, $\langle\zeta^2\rangle=0.153^{+0.014}_{-0.016}$, and $\langle\zeta^4\rangle=0.055^{+0.009}_{-0.009}$. The heavy-light momentum partition is controlled mainly by $\beta_{b,ud}$ and $m_b$, whereas the internal $u$--$d$ sharing is most sensitive to $\beta_{ud}$. The result provides a compact finite-mass benchmark for three-quark longitudinal structure, with distinct observables for heavy-light momentum sharing and correlations inside the light pair.
\end{abstract}

\maketitle

\section{Introduction}\label{sec:intro}

Light-front dynamics provides a relativistic description of hadrons in terms of longitudinal momentum fractions, transverse momenta, and constituent helicities. Its simple boost properties make light-front wave functions particularly useful for overlap representations of hadronic observables and for resolving the internal momentum structure of relativistic bound states \cite{Lepage:1980fj,Brodsky:1997de}. Relativistic constituent formulations are closely connected with the Bakamjian-Thomas construction and front-form Hamiltonian dynamics \cite{Bakamjian:1953kh,Keister:1991sb}. For a genuine three-quark baryon, two independent longitudinal fractions are required, with $x_u+x_d+x_b=1$. A two-variable distribution is therefore the minimal longitudinal description that keeps the momentum sharing of all three valence constituents explicit.

In QCD, baryon light-cone distribution amplitudes (LCDAs) are renormalized matrix elements of gauge-invariant three-quark light-ray operators. Their twist structure, scale dependence, conformal expansion, and recent lattice and LaMET studies have been investigated extensively for light baryons \cite{Braun:2000kw,QCDSF:2008qtn,Deng:2023csv,Han:2024ucv,LatticeParton:2024vck,Shi:2026vok}. Heavy baryons introduce an additional hierarchy because the heavy-quark momentum is parametrically different from the momenta carried by the light degrees of freedom. For the $\Lambda_b$, the heavy-quark limit leads naturally to a heavy-quark effective theory (HQET) description in terms of the light-quark energies $\omega_u$ and $\omega_d$. The complete three-quark LCDA basis, leading-twist evolution, and QCD-sum-rule models were developed in Refs.~\cite{Ball:2008fw,Ali:2012pn,Bell:2013tfa}, and the perturbatively generated radiative tail has recently been analyzed explicitly \cite{Feldmann:2025dcs}. These nonperturbative functions enter light-cone and soft-collinear effective theory sum rules for $\Lambda_b$ transition form factors and rare decays \cite{Wang:2009hra,Wang:2015ndk,Miao:2022bga,Huang:2022lfr,Khodjamirian:2023wol,Huang:2024oik,Lu:2025gjt}, and also serve as essential inputs in perturbative-QCD analyses of $\Lambda_b$ transition form factors and hadronic decays \cite{Han:2022srw,Han:2024kgz,Han:2025tvc}.

The two-variable nature of the light sector is physically important. Besides the total longitudinal momentum carried by the $ud$ pair, one may ask how that momentum is shared internally between the two light quarks. The two pieces of information are distinct: the first characterizes the heavy-light partition, whereas the second probes correlations inside the light pair. A quark-diquark reduction is efficient for many transition calculations, but by construction it compresses the two light constituents into one effective degree of freedom and therefore does not retain this internal sharing as an independent longitudinal variable. A three-quark treatment is required when the aim is to resolve both scales simultaneously.

A finite-mass constituent light-front wave function, however, describes an object different from an HQET LCDA. It is naturally expressed in dimensionless fractions $x_i$ and transverse momenta and does not carry the ultraviolet operator definition or the renormalization-scale dependence of $\psi_2(\omega_u,\omega_d;\mu)$. Even the relation between field-theoretic QCD and boosted-HQET LCDAs involves a nontrivial perturbative matching kernel \cite{Shi:2026mjb}. A constituent light-front quark model (LFQM) projection should therefore be interpreted as a model-scale valence quantity rather than identified directly with a renormalized HQET LCDA.

Relativistic light-front quark models provide a useful framework for finite-mass hadron structure and transition amplitudes \cite{Schlumpf:1992vq,Jaus:1999zv,Cheng:2003sm,Geng:2021nkl,Deng:2023qaf}. A fully three-quark covariant construction for baryons was developed by Zhao et al.~\cite{Zhao:2023yuk}, in which the relative motion between the heavy quark and the light pair and the internal motion of the two light quarks are described by two independent Gaussian parameters. This separation of momentum scales makes the framework well suited to a two-variable longitudinal analysis. The same three-quark setup has subsequently been applied to triply heavy baryons \cite{Lu:2024bqw}, while complementary BLFQ calculations of the $\Lambda_b$ and $\Sigma_b$ exhibit the expected dominance of the heavy-quark longitudinal momentum fraction \cite{Meng:2025sys}.

In this work we use the finite-mass three-quark wave function of Ref.~\cite{Zhao:2023yuk} to determine the normalized $C\gamma_5\gamma^+$ longitudinal projection $\Psi_2^{\rm LF}(x_u,x_d)$. The $u$ and $d$ quarks remain explicit constituents, and both independent transverse momenta are integrated without introducing an elementary diquark. The distribution has support on $x_u>0$, $x_d>0$, and $x_u+x_d<1$, and retains both the heavy--light and internal-$ud$ momentum scales of the parent wave function. We analyze its two-dimensional shape, fixed-$x_d$ slices, single-parton marginals, moments of the total light-pair momentum $s=x_u+x_d$, and moments of the internal-sharing variable $\zeta=(x_u-x_d)/(x_u+x_d)$. We also quantify the sensitivity of these observables to the constituent masses and Gaussian parameters. The subscript ``2'' labels the $C\gamma_5\gamma^+$ projection and does not denote an independently renormalized twist classification.

The paper is organized as follows. Section~\ref{sec:lfstate} summarizes the three-quark light-front state. Section~\ref{sec:def} defines the longitudinal projection and derives its overlap representation. Section~\ref{sec:results} presents the numerical results, model-parameter sensitivity, and physical interpretation. Section~\ref{sec:conclusion} summarizes the main conclusions.

\section{Three-quark light-front state}\label{sec:lfstate}

\subsection{Kinematics}

The $\Lambda_b$ baryon is treated as a three-quark state composed of a heavy $b$ quark and a light $ud$ pair. We adopt the light-front convention $p^\pm=p^0\pm p^3$ and denote the momentum of each constituent quark by $\widetilde p_i=(p_i^+,\bm p_{i\perp})$, while the total baryon momentum is denoted by $\widetilde P=(P^+,\bm P_\perp)$, with $i\in\{b,d,u\}$. The intrinsic variables are defined by
\begin{equation}
 p_i^+=x_iP^+,\qquad
 \bm p_{i\perp}=x_i\bm P_\perp+\bm k_{i\perp},
 \label{eq:intrinsic}
\end{equation}
subject to
\begin{equation}
 \sum_i x_i=1,\qquad
 \sum_i\bm k_{i\perp}=0.
 \label{eq:constraints}
\end{equation}
The free invariant mass of the three-quark system is
\begin{equation}
 M_0^2=\sum_i\frac{m_i^2+\bm k_{i\perp}^2}{x_i},
 \label{eq:M0}
\end{equation}
and the internal energies and longitudinal momenta in the $\Pbar$ rest frame are
\begin{equation}
 e_i=\frac{x_iM_0}{2}+\frac{m_i^2+\bm k_{i\perp}^2}{2x_iM_0},\qquad
 k_{iz}=\frac{x_iM_0}{2}-\frac{m_i^2+\bm k_{i\perp}^2}{2x_iM_0}.
 \label{eq:kz}
\end{equation}
These definitions imply $e_i=\sqrt{m_i^2+\bm k_i^2}$ and $e_i+k_{iz}=x_iM_0$, where $\bm k_i=(\bm k_{i\perp},k_{iz})$ denotes the corresponding three-dimensional internal momentum.

We define
\begin{equation}
 \Pbar=p_b+p_d+p_u,\qquad \Pbar^2=M_0^2,
 \label{eq:Pbar}
\end{equation}
where $\Pbar$ is the total momentum of the free three-quark system, whereas $P$ is the physical baryon momentum satisfying $P^2=M_{\Lambda_b}^2$. The two momenta share the same kinematic light-front components, so that $\widetilde{\Pbar}=\widetilde P$, but their minus components differ in general because $M_0$ is the free invariant mass rather than the physical baryon mass.

The three-quark Fock state is written as \cite{Zhao:2023yuk}
\begin{equation}
 |\Lb(P,S,S_z)\rangle=\int\prod_i\{d^3\widetilde p_i\}\,2(2\pi)^3
 \delta^3\!\left(\widetilde P-\sum_i\widetilde p_i\right)\frac{1}{\sqrt{P^+}}
 \sum_{\lambda_b,\lambda_d,\lambda_u}
 \Psi^{SS_z}(\widetilde p_b,\widetilde p_d,\widetilde p_u;
 \lambda_b,\lambda_d,\lambda_u)
 \frac{\epsilon^{\alpha\beta\gamma}}{\sqrt6}
 |b_\alpha d_\beta u_\gamma\rangle,
 \label{eq:fock}
\end{equation}
where $\alpha,\beta,\gamma$ are color indices and
\begin{equation}
 \{d^3\widetilde p\}=\frac{dp^+\,d^2\bm p_\perp}{2(2\pi)^3}.
 \label{eq:lfmeasure}
\end{equation}

\subsection{Spin and momentum wave functions}

For the scalar $ud$ channel, the covariant spin-momentum wave function is
\begin{equation}
 \Psi^{1/2,S_z}_0=A_0\,
 \bar u(p_u,\lambda_u)(\slashed{\Pbar}+M_0)(-\gamma_5)C\bar u^T(p_d,\lambda_d)\,
 \bar u(p_b,\lambda_b)u(\Pbar,S_z)\,\Phi(x_i,\bm k_{i\perp}),
 \label{eq:spinwf}
\end{equation}
with
\begin{equation}
 A_0=\frac{1}{4\sqrt{M_0^3(e_b+m_b)(e_d+m_d)(e_u+m_u)}}.
 \label{eq:A0}
\end{equation}
The momentum wave function is
\begin{equation}
\Phi=
\sqrt{\frac{e_be_de_u}{x_bx_dx_uM_0}}
\varphi(\bm{k}_b;\beta_{b,ud})
\varphi\!\left(\frac{\bm{k}_d-\bm{k}_u}{2};\beta_{ud}\right).
 \label{eq:Phi}
\end{equation}
where
\begin{equation}
 \varphi(\bm k;\beta)=4\left(\frac{\pi}{\beta^2}\right)^{3/4}
 \exp\!\left[-\frac{\bm k_\perp^2+k_z^2}{2\beta^2}\right].
 \label{eq:gaussian}
\end{equation}
Here the subscript \(ud\) denotes the two-light-quark subsystem and does not imply an elementary diquark degree of freedom. The parameter \(\beta_{b,ud}\) controls the relative motion of the \(b\) quark with respect to this subsystem, while \(\beta_{ud}\) describes the internal relative motion of the \(u\) and \(d\) quarks.

\section{Definition and overlap representation}\label{sec:def}

\subsection{Two-variable light-front distribution}

Let $n^\mu$ be a lightlike direction with $\slashed n=\gamma^+$ in the adopted convention. In light-cone gauge the straight Wilson lines along $n$ reduce to unity. We consider
\begin{equation}
 {\cal O}_2(t_1,t_2)=\epsilon^{abc}
 \left[u_a^T(t_1n)C\gamma_5\slashed n\,d_b(t_2n)\right]b_c(0).
 \label{eq:operator}
\end{equation}
The valence overlap defines the two-variable longitudinal distribution through
\begin{equation}
 {\cal M}_2(t_1,t_2)\propto
 \int_0^1dx_u\int_0^{1-x_u}dx_d\,
 e^{-iP^+(t_1x_u+t_2x_d)}\Psi_2^{\rm LF}(x_u,x_d),
 \label{eq:fourier}
\end{equation}
with the shape normalization
\begin{equation}
 \int_0^1dx_u\int_0^{1-x_u}dx_d\,
 \Psi_2^{\rm LF}(x_u,x_d)=1.
 \label{eq:norm}
\end{equation}
This normalization isolates the longitudinal shape from the local-current coupling. Determining an absolute renormalized coupling would require specifying the renormalization scheme and scale, together with the corresponding matching prescription, which is beyond the present model calculation.

\subsection{Color, helicity sums, and Dirac structures}

Contracting the color tensor with the normalized baryon color wave function gives $\sqrt6$. The constituent helicity sums are performed with the standard completeness relation $\sum_\lambda u(p,\lambda)\bar u(p,\lambda)=\slashed p+m$. The light-quark part becomes
\begin{equation}
 {\cal T}_2=\Tr\!\big[C\gamma_5\slashed n(\slashed p_u+m_u)
 (\slashed{\Pbar}+M_0)(-\gamma_5)C(\slashed p_d+m_d)^T\big].
 \label{eq:trace0}
\end{equation}
Using $C\gamma_\mu^TC^{-1}=-\gamma_\mu$ and $C^T=-C$, one obtains
\begin{equation}
 {\cal T}_2=4\big\{(n\!\cdot\!p_u)(\Pbar\!\cdot\!p_d+M_0m_d)
 +(n\!\cdot\!\Pbar)(m_um_d-p_u\!\cdot\!p_d)
 +(n\!\cdot\!p_d)(p_u\!\cdot\!\Pbar+m_uM_0)\big\}.
 \label{eq:tracecov}
\end{equation}
With $n\cdot p_i=x_iP^+$ and $n\cdot\Pbar=P^+$, the reduced trace is
\begin{equation}
 \widehat{\cal T}_2\equiv\frac{{\cal T}_2}{4P^+}
 =M_0x_u(e_d+m_d)+M_0x_d(e_u+m_u)+m_um_d-p_u\cdot p_d,
 \label{eq:tracereduced}
\end{equation}
where
\begin{equation}
 p_u\cdot p_d=\frac12\!\left[
 \frac{x_u}{x_d}(m_d^2+k_{d\perp}^2)+\frac{x_d}{x_u}(m_u^2+k_{u\perp}^2)\right]
 -\bm k_{u\perp}\cdot\bm k_{d\perp}.
 \label{eq:pupd}
\end{equation}

The heavy-spinor line is projected with $\gamma^+$,
\begin{equation}
 {\cal H}_b=\frac{\Tr[\gamma^+(\slashed p_b+m_b)(\slashed{\Pbar}+M_0)]}
 {\Tr[\gamma^+(\slashed P+M_{\Lb})]}
 =m_b+x_bM_0.
 \label{eq:heavy}
\end{equation}
The good-component projection avoids additional dependence on the mismatch $\Pbar\ne P$.

\subsection{Final transverse-momentum integral}

Combining Eqs.~\eqref{eq:A0}, \eqref{eq:Phi}, \eqref{eq:tracereduced}, and \eqref{eq:heavy}, we define
\begin{equation}
 \rho_2(x_u,x_d)=\int\frac{d^2\bm k_{u\perp}}{2(2\pi)^3}
 \frac{d^2\bm k_{d\perp}}{2(2\pi)^3}
 \frac{\sqrt6\,A_0\Phi}{\sqrt{x_bx_ux_d}}
 (m_b+x_bM_0)\widehat{\cal T}_2,\qquad x_b=1-x_u-x_d,
 \label{eq:rho}
\end{equation}
and
\begin{equation}
 \Psi_2^{\rm LF}(x_u,x_d)=
 \frac{\rho_2(x_u,x_d)}
 {\int_0^1dy_u\int_0^{1-y_u}dy_d\,\rho_2(y_u,y_d)}.
 \label{eq:finalpsi}
\end{equation}
Overall convention-dependent constants cancel in Eq.~\eqref{eq:finalpsi}; the shape is fixed by the three-quark kinematics, spin structure, constituent masses, and two Gaussian parameters.

For the numerical integration we remove one overall transverse azimuth and write
\begin{equation}
 d^2\bm k_{u\perp}d^2\bm k_{d\perp}
 =2\pi\,k_udk_u\,k_ddk_d\,d\theta,
\end{equation}
where $\theta$ is the relative azimuth. The radial intervals are mapped to finite intervals through
\begin{equation}
 k=\kappa\frac{t}{1-t},\qquad 0<t<1,
 \label{eq:kmap}
\end{equation}
which covers $k\in[0,\infty)$ without a hard transverse cutoff.

The Dirac reduction, good-component heavy-quark identity, transverse measure, and numerical normalization were checked independently of the production calculation. The essential cross-checks are summarized in the Appendix~\ref{app:checks}.

\section{Numerical results and discussion}\label{sec:results}

\subsection{Input and numerical implementation}

The central constituent masses and Gaussian parameters follow Ref.~\cite{Zhao:2023yuk}. We use $m_q=m_u=m_d=0.250~\mathrm{GeV}$, $m_b=4.800~\mathrm{GeV}$, $\beta_{b,ud}=0.630~\mathrm{GeV}$, and $\beta_{ud}=0.270~\mathrm{GeV}$. The parameter variations quoted in Ref.~\cite{Zhao:2023yuk} are retained as model ranges. Because statistical uncertainties are not assigned to the constituent masses there, the mass windows used below are treated only as sensitivity probes. The inputs are summarized in Table~\ref{tab:setup}.

\begin{table}[h]
\caption{Central model inputs and adopted variation ranges (in GeV).}
\label{tab:setup}
\renewcommand{\arraystretch}{1}
\renewcommand{\tabcolsep}{1pc}
\centering
\begin{tabular}{lccccc}
\hline\hline
Input & $m_q$  & $m_b$  & $M_{\Lambda_b}$ &
$\beta_{b,ud}$  & $\beta_{ud}$  \\
\hline
Value &
$0.250\pm0.020$ &
$4.800\pm0.200$ &
$5.620$ &
$0.630\pm0.050$ &
$0.270\pm0.030$ \\
\hline\hline
\end{tabular}
\end{table}

\begin{table*}[t]
\caption{Selected moments of the normalized two-variable distribution.}
\label{tab:moments}
\centering
\small
\renewcommand{\arraystretch}{1.2}
\renewcommand{\tabcolsep}{1pc}
\begin{tabular}{ccccccccc}
\hline\hline
$\langle x_u\rangle=\langle x_d\rangle$
& $\langle x_b\rangle$
& $\langle s\rangle$
& $\langle s^2\rangle$
& $\langle\zeta^2\rangle$
& $\langle\zeta^4\rangle$
& $\langle s^2\zeta^2\rangle$
& $\langle s^3\zeta^3\rangle$
& $\langle s^2\zeta^2\rangle/
        (\langle s^2\rangle\langle\zeta^2\rangle)$
\\
\hline
0.130
& 0.739
& 0.261
& 0.078
& 0.153
& 0.055
& 0.0105
& 0
& 0.877
\\
\hline\hline
\end{tabular}
\end{table*}
The transverse integrations are evaluated with deterministic Gauss--Legendre quadrature after the mapping in Eq.~\eqref{eq:kmap}. The triangular longitudinal domain is parametrized by
\begin{equation}
 s=x_u+x_d,\qquad
 u=\frac{x_u}{x_u+x_d},\qquad
 x_u=su,\qquad
 x_d=s(1-u),\qquad
 dx_u\,dx_d=s\,ds\,du.
 \label{eq:trianglemap}
\end{equation}
We take $\kappa=0.800~\mathrm{GeV}$ as a numerical mapping scale. The light-quark masses are varied simultaneously to preserve isospin symmetry.

\begin{figure}[!t]
 \centering
 %\hspace*{0.018\textwidth}%
 \includegraphics[width=0.7\columnwidth]{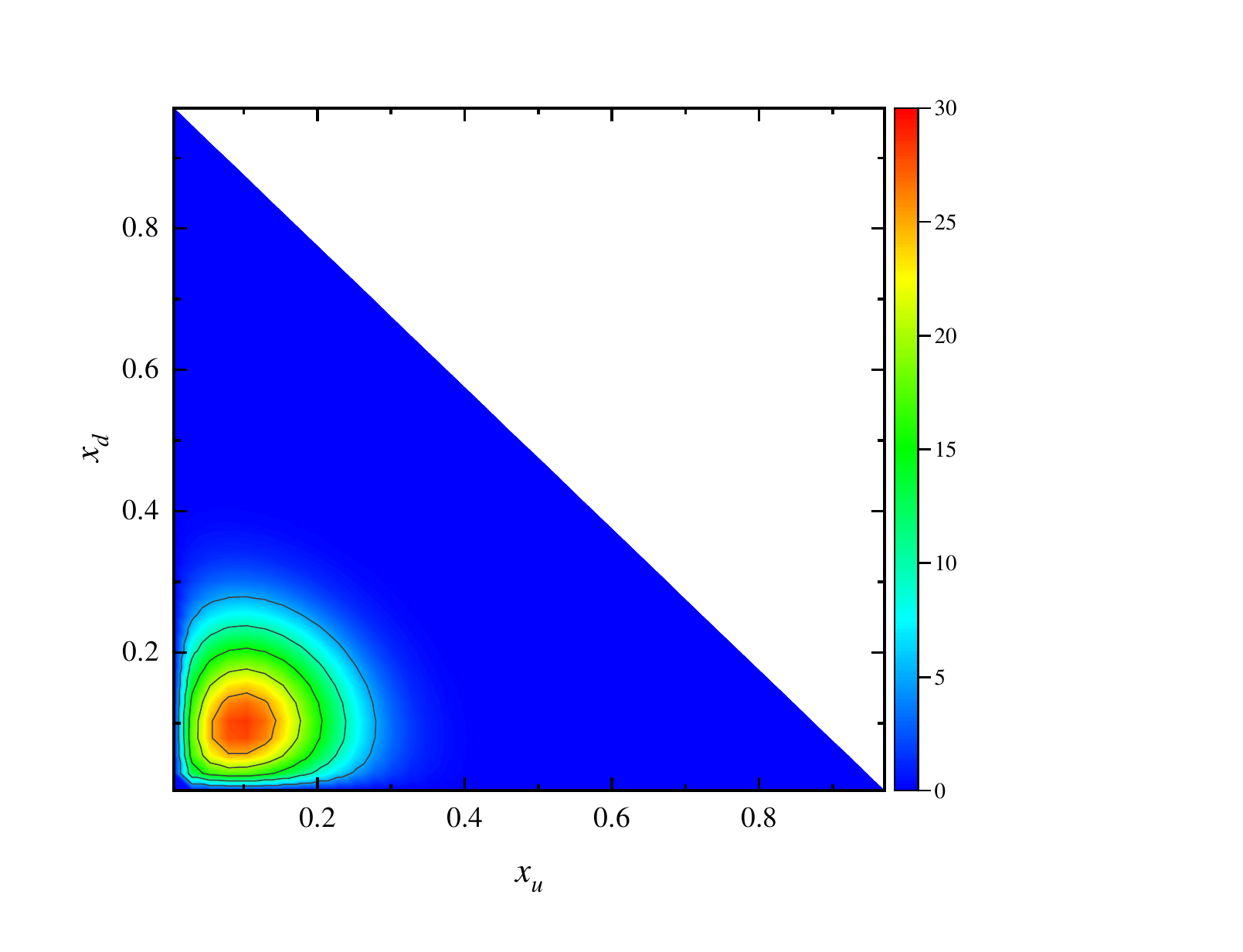}
 \caption{Normalized two-variable distribution $\Psi_2^{\rm LF}(x_u,x_d)$ for the central parameter set. The distribution is defined on the physical triangle $x_u>0$, $x_d>0$, and $x_u+x_d<1$, and is concentrated in the small-$x_u$, small-$x_d$ region.}
 \label{fig:density}
\end{figure}

\begin{figure}[!t]
 \centering
 \includegraphics[width=0.7\columnwidth]{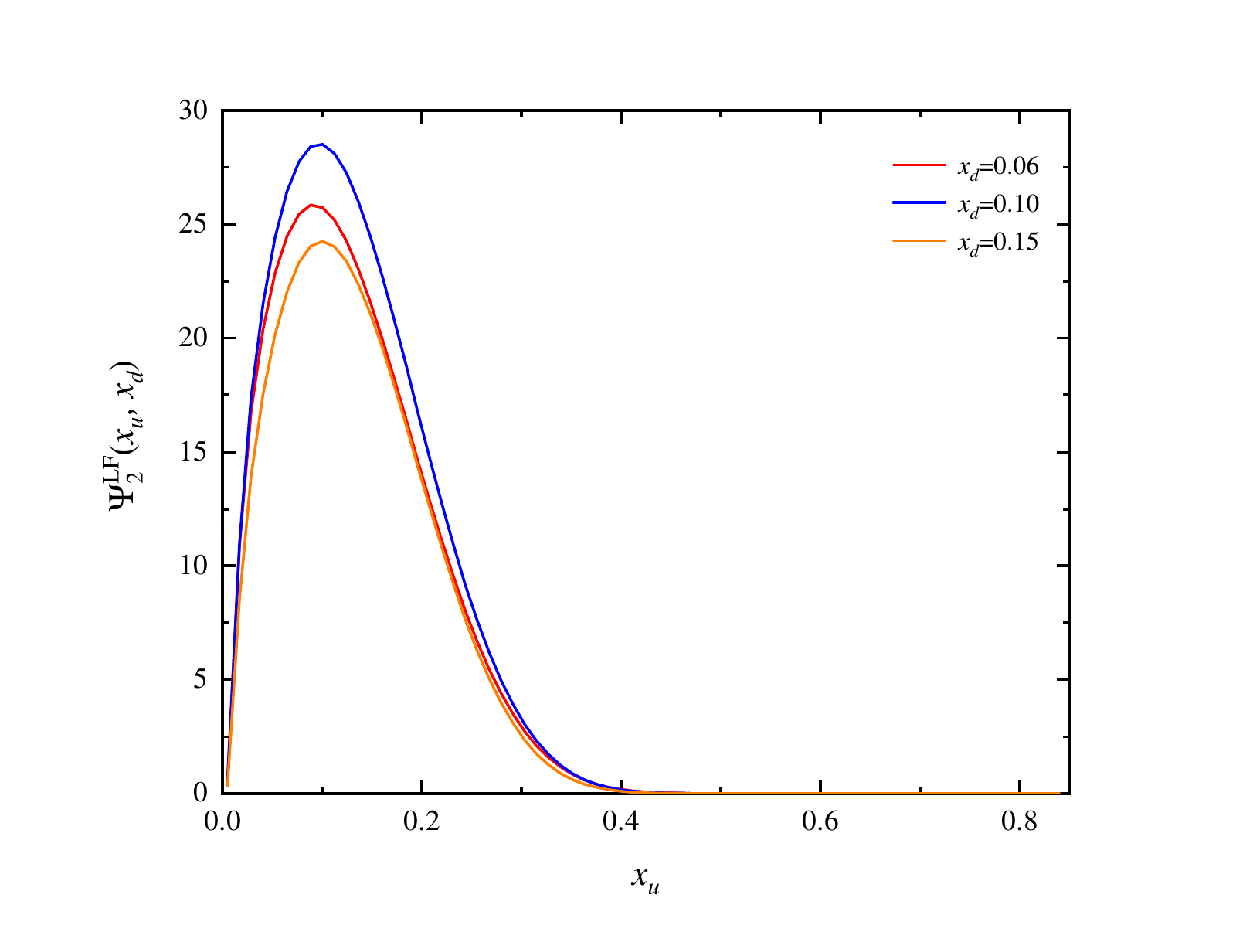}
 \caption{Fixed-$x_d$ slices of $\Psi_2^{\rm LF}(x_u,x_d)$ for $x_d=0.060$, $0.100$, and $0.150$, shown over the common interval $0.005\leq x_u\leq0.840$. This range lies entirely within the physical domain $x_u<1-x_d$ for all three slices. The distribution is strongly suppressed at large $x_u$.}
 \label{fig:slices}
\end{figure}

\begin{figure}[!t]
 \centering
 \includegraphics[width=0.7\columnwidth]{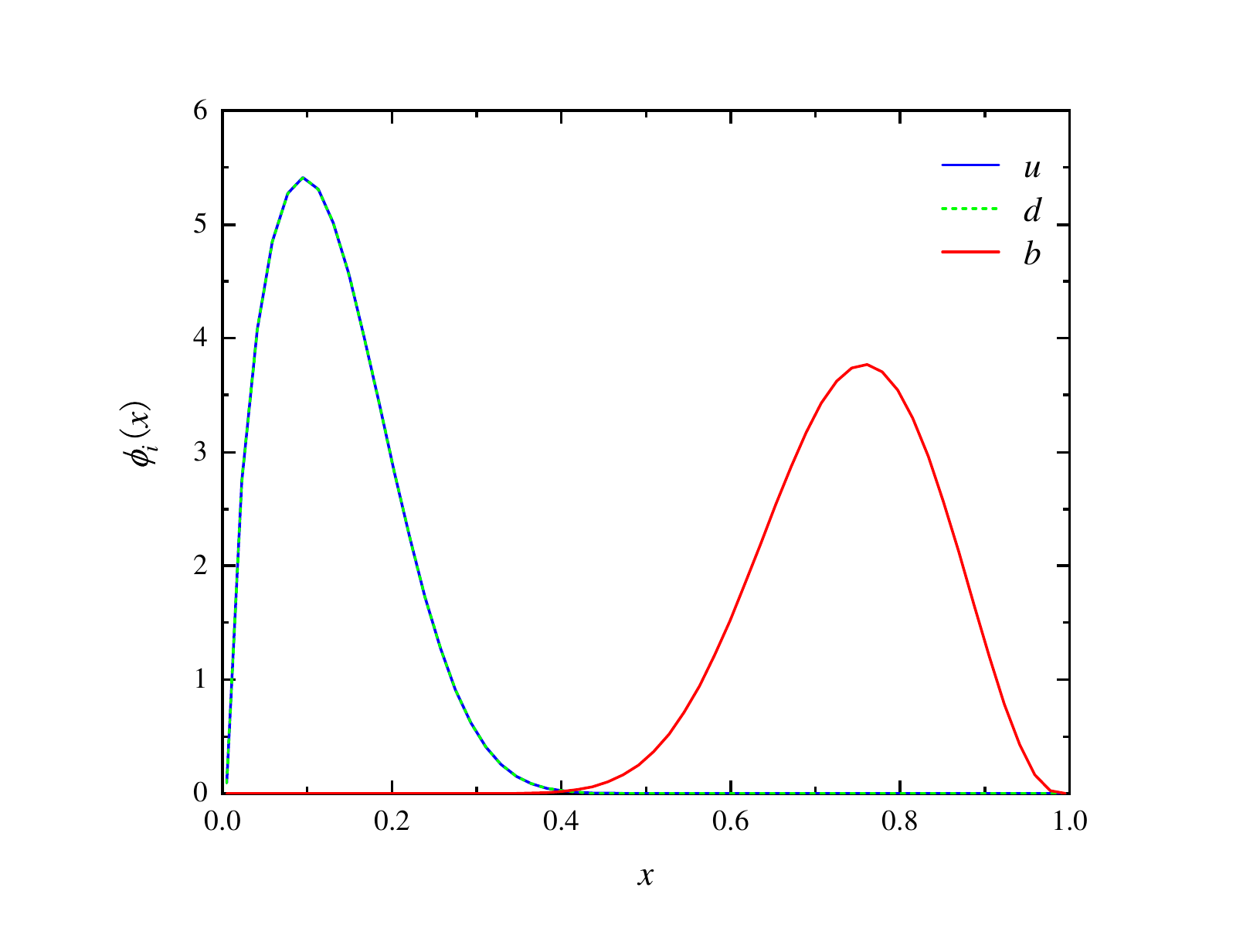}
 \caption{Single-parton longitudinal marginals for the central parameter set. The $u$ and $d$ distributions coincide in the isospin limit, whereas the $b$-quark distribution is concentrated at substantially larger momentum fraction.}
 \label{fig:partonmarg}
\end{figure}

\subsection{Two-variable distribution and longitudinal moments}

Figs.~\ref{fig:density}--\ref{fig:partonmarg} summarize the central longitudinal structure. Fig.~\ref{fig:density} shows that the two-variable distribution is localized well inside the physical triangle and is largest when both light-quark fractions are small. Fig.~\ref{fig:slices} confirms the rapid suppression toward large $x_u$ at fixed $x_d$, while Fig.~\ref{fig:partonmarg} displays the complementary hierarchy between the light- and heavy-quark single-parton marginals. To characterize the internal sharing independently of the total light-pair momentum, we introduce
\begin{equation}
 \zeta=2u-1=\frac{x_u-x_d}{x_u+x_d}.
 \label{eq:szeta}
\end{equation}
For any function $f(x_u,x_d)$,
\begin{equation}
 \langle f\rangle=\int_0^1dx_u\int_0^{1-x_u}dx_d\,
 f(x_u,x_d)\Psi_2^{\rm LF}(x_u,x_d).
 \label{eq:momdef}
\end{equation}

The central moments are listed in Table~\ref{tab:moments}. The average momentum fraction carried by the two light quarks is $\langle s\rangle=0.261$, while the $b$ quark carries about $73.9\%$ of the longitudinal momentum. In the isospin-symmetric limit adopted here, the distribution satisfies $\Psi_2^{\rm LF}(x_u,x_d)=\Psi_2^{\rm LF}(x_d,x_u)$. Since $s=x_u+x_d$ is invariant whereas $\zeta=(x_u-x_d)/(x_u+x_d)$ changes sign under $x_u\leftrightarrow x_d$, all moments containing an odd power of $\zeta$ vanish. In particular, $\langle x_u-x_d\rangle=0$ and $\langle s^3\zeta^3\rangle=0$. For the even mixed moment, we obtain $\langle s^2\zeta^2\rangle=0.0105$ and $\langle s^2\zeta^2\rangle/(\langle s^2\rangle\langle\zeta^2\rangle)=0.877$. The deviation of this ratio from unity shows that the dependence on the total light-pair momentum and its internal sharing does not factorize exactly. In addition, $\langle\zeta^2\rangle=0.153$ and $\langle\zeta^4\rangle=0.055$, both below the values $1/3$ and $1/5$ corresponding to a uniform distribution in $u$. Thus the light pair preferentially occupies configurations with $x_u\simeq x_d$, while its total momentum $s$ remains broadly distributed.

\subsection{Model-parameter sensitivity}

We estimate model sensitivity by varying $m_q$, $m_b$, $\beta_{b,ud}$, and $\beta_{ud}$ independently over the ranges in Table~\ref{tab:setup}. For an observable $O$, the upward and downward shifts from varied parameter $a$ are defined as
\begin{equation}
 \delta_a^+=\max\bigl(0,O_a^--O_0,O_a^+-O_0\bigr),\qquad
 \delta_a^-=\max\bigl(0,O_0-O_a^-,O_0-O_a^+\bigr).
 \label{eq:parame}
\end{equation}
For each parameter \(a\), \(\delta_a^+\) and \(\delta_a^-\) define the upward and downward one-sided envelopes of its two variations. Assuming independent parameter variations, these one-sided contributions are combined separately in quadrature,
\begin{equation}
 \Delta O_\pm=\left[\sum_a(\delta_a^\pm)^2\right]^{1/2}.
 \label{eq:paramerror}
\end{equation}
The resulting \(\Delta O^+\) and \(\Delta O^-\) are the final upward and downward model-sensitivity estimates, without any further quadrature combination, and the corresponding ranges are summarized in Table~\ref{tab:uncertainties}.

\begin{table*}[t]
\caption{Individual one-sided parameter shifts and the combined model sensitivities of selected longitudinal observables. For each varied parameter $a$, the entries denote the upward and downward envelopes $^{+\delta_a^+}_{-\delta_a^-}$ defined in Eq.~(\ref{eq:parame}). The last row gives the quadrature combination of the four parameter contributions according to Eq.~(\ref{eq:paramerror}).}
\label{tab:uncertainties}
\centering
\begin{ruledtabular}
\begin{tabular}{lcccccc}
 & $\langle x_u\rangle=\langle x_d\rangle$
 & $\langle x_b\rangle$
 & $\langle s\rangle$
 & $\langle s^2\rangle$
 & $\langle\zeta^2\rangle$
 & $\langle\zeta^4\rangle$ \\
\hline

Central
& 0.130 & 0.739 & 0.261 & 0.078 & 0.153 & 0.055 \\

$m_q$
& ${}^{+<0.001}_{-<0.001}$
& ${}^{+0.001}_{-0.001}$
& ${}^{+0.001}_{-0.001}$
& ${}^{+<0.001}_{-<0.001}$
& ${}^{+0.004}_{-0.004}$
& ${}^{+0.003}_{-0.003}$ \\

$m_b$
& ${}^{+0.004}_{-0.004}$
& ${}^{+0.007}_{-0.008}$
& ${}^{+0.008}_{-0.007}$
& ${}^{+0.005}_{-0.004}$
& ${}^{+<0.001}_{-<0.001}$
& ${}^{+<0.001}_{-<0.001}$ \\

$\beta_{b,ud}$
& ${}^{+0.005}_{-0.005}$
& ${}^{+0.010}_{-0.009}$
& ${}^{+0.009}_{-0.010}$
& ${}^{+0.006}_{-0.006}$
& ${}^{+0.006}_{-0.005}$
& ${}^{+0.003}_{-0.003}$ \\

$\beta_{ud}$
& ${}^{+0.003}_{-0.003}$
& ${}^{+0.006}_{-0.006}$
& ${}^{+0.006}_{-0.006}$
& ${}^{+0.003}_{-0.003}$
& ${}^{+0.012}_{-0.014}$
& ${}^{+0.008}_{-0.008}$ \\

Combined $\Delta O$
& ${}^{+0.007}_{-0.007}$
& ${}^{+0.014}_{-0.014}$
& ${}^{+0.014}_{-0.014}$
& ${}^{+0.008}_{-0.008}$
& ${}^{+0.014}_{-0.016}$
& ${}^{+0.009}_{-0.009}$ \\

\end{tabular}
\end{ruledtabular}
\end{table*}

\begin{figure}[!t]
 \centering
 \begin{minipage}[t]{0.485\textwidth}
  \centering
  \includegraphics[width=\linewidth]{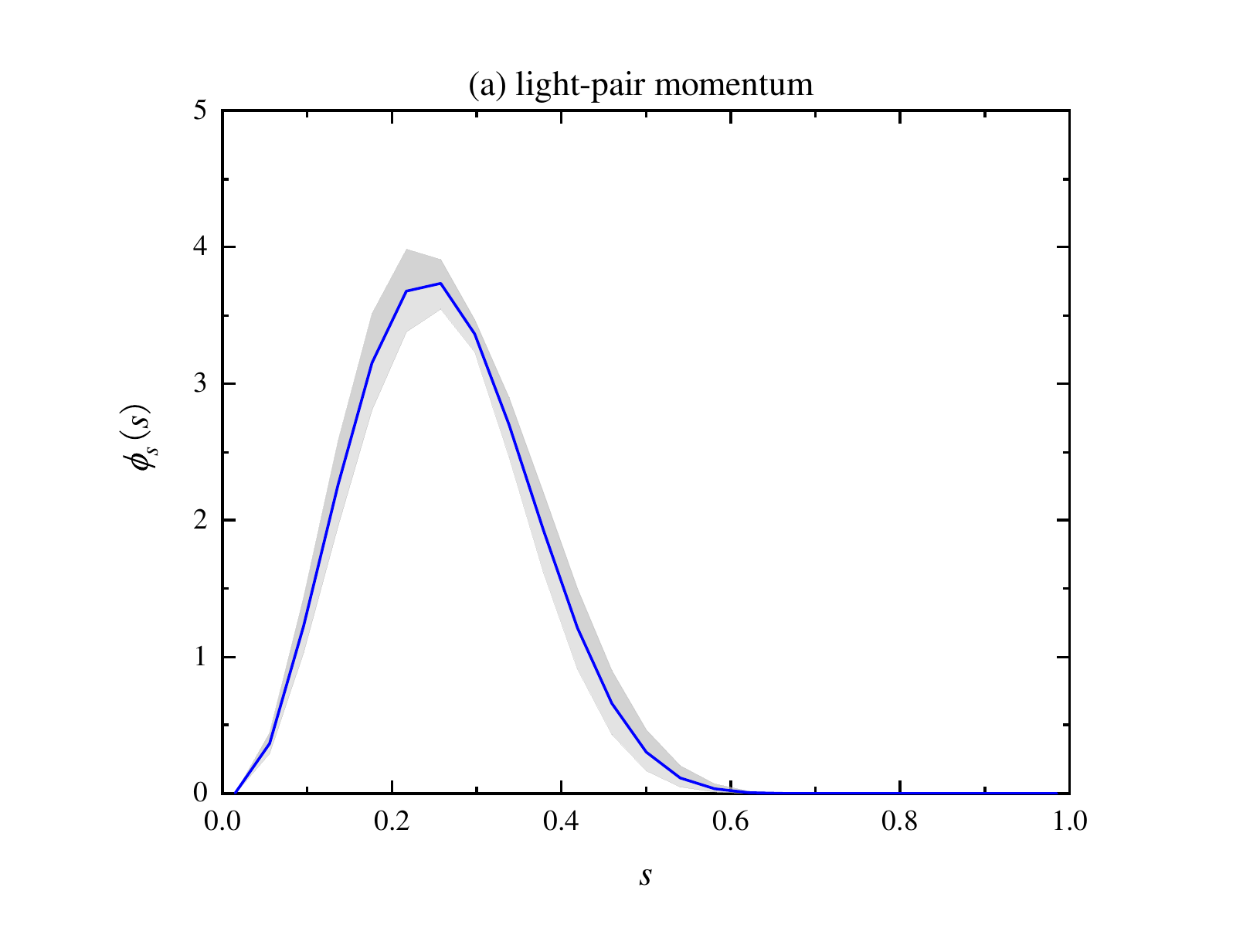}
 \end{minipage}
 \hfill
 \begin{minipage}[t]{0.485\textwidth}
  \centering
  \includegraphics[width=\linewidth]{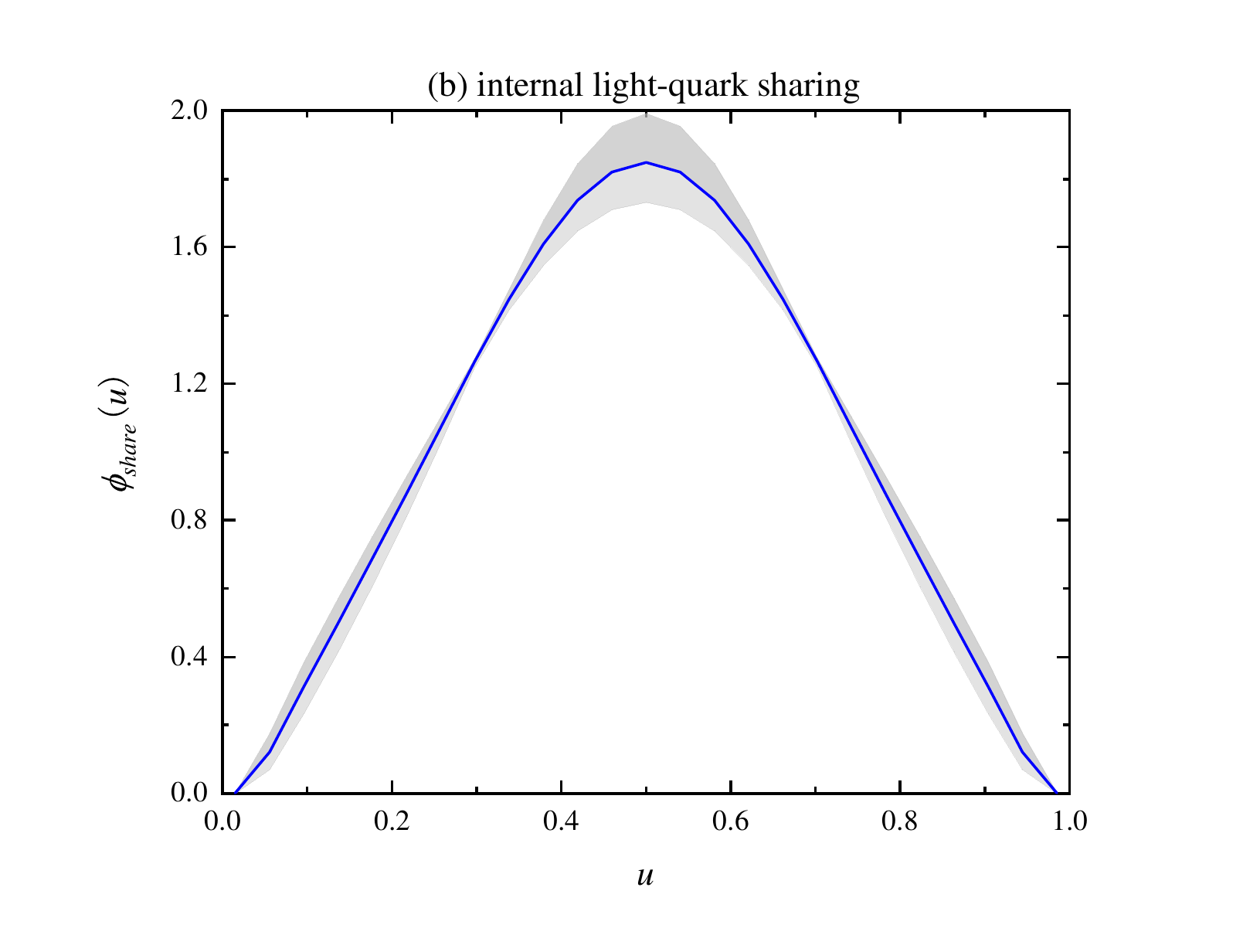}
 \end{minipage}
 \caption{Pointwise model-sensitivity bands for (a) the normalized total light-pair momentum distribution $\phi_s(s)$ and (b) the internal-sharing distribution $\phi_{\rm share}(u)$. The solid curves show the central results, while the shaded regions represent the variations induced by the adopted parameter ranges.}
 \label{fig:uncertaintybands}
\end{figure}

Using the variables $s$ and $u$ in Eq.~\eqref{eq:trianglemap}, the normalized one-dimensional projections are
\begin{equation}
 \phi_s(s)=s\int_0^1du\,\Psi_2^{\rm LF}\!\bigl(su,s(1-u)\bigr),\qquad
 \phi_{\rm share}(u)=\int_0^1ds\,s\,\Psi_2^{\rm LF}\!\bigl(su,s(1-u)\bigr).
 \label{eq:marginalsSU}
\end{equation}
Fig.~\ref{fig:uncertaintybands} shows the corresponding central shapes and pointwise model-sensitivity bands. The distribution $\phi_s(s)$ is concentrated in the low-$s$ region, with $\langle s\rangle=0.261$, whereas $\phi_{\rm share}(u)$ is symmetric about $u=1/2$ and reaches its maximum at equal light-quark sharing. The narrowing of the $\phi_{\rm share}(u)$ sensitivity band near $u\simeq0.3$ and $0.7$ results from approximate crossing points between the central and separately normalized parameter-varied distributions. The symmetric occurrence of these narrow regions reflects the symmetry of the light quark pair.

\begin{figure}[!t]
 \centering
 \includegraphics[width=0.7\columnwidth]{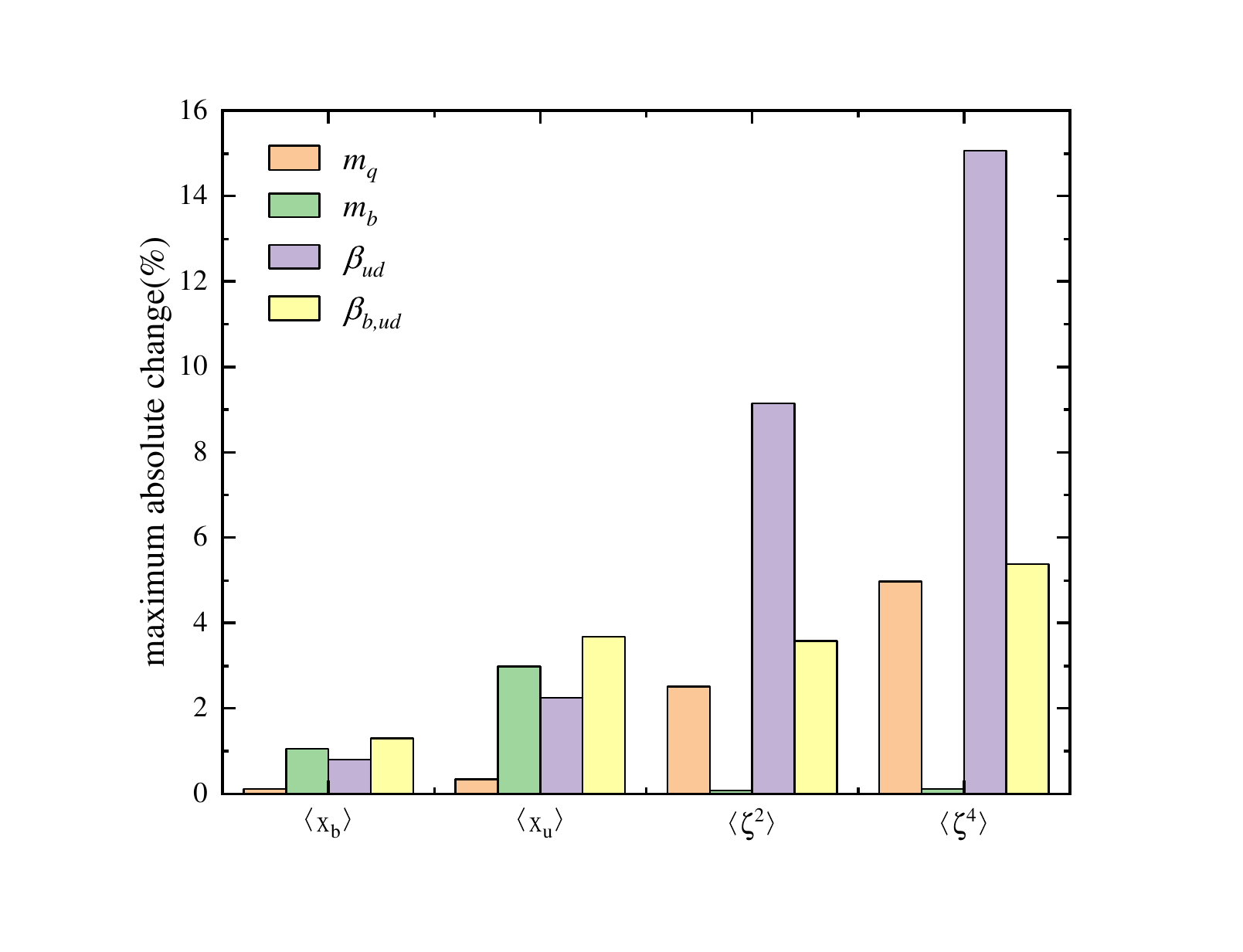}
\caption{Maximum absolute relative changes of selected longitudinal observables under the adopted variations of \(m_q\), \(m_b\), \(\beta_{ud}\), and \(\beta_{b,ud}\). For each parameter \(a\), the plotted value is the larger of the upward and downward one-sided relative shifts, expressed as a percentage, with \(\delta_a^\pm\) defined in Eq.~(\ref{eq:parame}).}
 \label{fig:sensitivityaaa}
\end{figure}

The parameter hierarchy is summarized in Fig.~\ref{fig:sensitivityaaa}. The plotted percentages represent the larger of the upward and downward one-sided shifts for each individual parameter, while the total sensitivities in Table~III are obtained by combining all four parameter contributions according to Eq.~(28). For \(\langle x_u\rangle\), the largest one-parameter shifts are induced by \(\beta_{b,ud}\) (\(3.7\%\)) and \(m_b\) (\(3.0\%\)), followed by \(\beta_{ud}\) (\(2.2\%\)), while the effect of \(m_q\) remains below \(0.4\%\). For the internal-sharing moments, \(\beta_{ud}\) changes \(\langle\zeta^2\rangle\) by up to \(9.2\%\) and \(\langle\zeta^4\rangle\) by up to \(15.1\%\), whereas the corresponding effect of \(m_b\) is below \(0.1\%\). This separation shows that \(\beta_{b,ud}\) primarily controls the heavy--light momentum partition, whereas \(\beta_{ud}\) governs the internal \(u\)--\(d\) sharing.

\subsection{Longitudinal momentum structure}

The resulting longitudinal structure exhibits the expected hierarchy of a finite-mass heavy--light baryon. The $b$ quark carries most of the longitudinal momentum, while the light pair remains concentrated at substantially smaller fractions. This behavior follows from the invariant mass in Eq.~\eqref{eq:M0}, since small $x_b$ enhances
$(m_b^2+\bm k_{b\perp}^2)/x_b$, whereas large light-quark fractions are suppressed by the Gaussian wave function. The finite-mass result therefore interpolates naturally between equal momentum sharing, $x_i=1/3$, and the static heavy-quark limit, $x_b=1$. In the two-dimensional distribution, the maximum occurs near $(x_u,x_d)=(0.104,0.104)$. Isospin symmetry favors nearly equal light-quark fractions, while the moments of $\zeta$ isolate the internal sharing from fluctuations in the total light-pair momentum. Together with the parameter study, these observables reveal a clear two-scale pattern: $\beta_{b,ud}$ primarily governs the heavy--light momentum partition, whereas $\beta_{ud}$ controls the internal $u$--$d$ sharing.

Qualitative support for the heavy-quark dominance is provided by the recent BLFQ study of the $\Lambda_b$ \cite{Meng:2025sys}. A pointwise comparison is nevertheless not meaningful, because BLFQ yields Hamiltonian eigenstates and evolved single-parton distribution functions (PDFs), whereas the present calculation isolates a normalized two-variable valence projection of a constituent light-front wave function. In particular, the present quantity retains the correlation between the two light fractions through $s$ and $\zeta$. An even more fundamental distinction arises in comparison with renormalized HQET LCDAs: QCD and HQET amplitudes possess explicit scale evolution and a perturbatively generated high-energy tail \cite{Ball:2008fw,Bell:2013tfa,Feldmann:2025dcs}, neither of which is contained in the Gaussian constituent wave function. Moreover, QCD and boosted-HQET LCDAs require perturbative matching already at the field-theory level \cite{Shi:2026mjb}.

Accordingly, $\Psi_2^{\rm LF}(x_u,x_d)$ should be interpreted as a model-scale valence longitudinal projection rather than as the renormalized HQET LCDA $\psi_2(\omega_u,\omega_d;\mu)$. Establishing a quantitative relation between the two would require a specified heavy-quark limit, matching of the current and normalization, and evolution to a common scale. The quoted uncertainty bands should likewise be viewed as model-sensitivity estimates rather than statistical or QCD errors, since they do not include variations of the radial ansatz, higher Fock components, gluonic degrees of freedom, or matching uncertainties. Within this scope, the two-variable distribution provides a compact characterization of the finite-mass three-quark longitudinal structure while retaining information that is absent in an elementary quark--diquark reduction.

\section{Conclusion}\label{sec:conclusion}

We have constructed a normalized two-variable longitudinal distribution of the $\Lambda_b=(ud)b$ from a finite-mass covariant three-quark light-front wave function. The $u$ and $d$ quarks remain explicit constituents, the scalar channel is selected with $C\gamma_5\gamma^+$, and both independent transverse momenta are integrated. The resulting $\Psi_2^{\rm LF}(x_u,x_d)$ is symmetric under $x_u\leftrightarrow x_d$, has support on the physical momentum-fraction triangle, and retains the two momentum scales of the parent wave function. For the central parameter set, $\langle x_u\rangle=\langle x_d\rangle=0.130$ and $\langle x_b\rangle=0.739$, while $\langle\zeta^2\rangle=0.153$ and $\langle\zeta^4\rangle=0.055$ show a clear preference for nearly equal internal $u$--$d$ sharing. The parameter analysis separates the two dynamical scales:
$\beta_{b,ud}$ and $m_b$ primarily determine the heavy--light momentum partition, whereas $\beta_{ud}$ dominates the internal light-pair sharing. The quantity obtained here is a model-scale valence light-front projection rather than the renormalized HQET LCDA $\psi_2(\omega_u,\omega_d;\mu)$. A quantitative connection between the two requires a separate matching and evolution analysis. Within its stated scope, $\Psi_2^{\rm LF}$ provides a two-variable benchmark for finite-mass three-quark longitudinal structure and for comparisons among alternative light-front wave functions.

\appendix
\section{Numerical and algebraic cross-checks}\label{app:checks}

The algebraic reduction in Sec.~\ref{sec:def} was checked independently of the production calculation. The $C$-eliminated light-quark trace was evaluated symbolically with FeynCalc 10.2.1 and reproduces Eq.~\eqref{eq:tracecov} exactly. An independent calculation using explicit $4\times4$ Dirac matrices with $C=i\gamma^2\gamma^0$ was applied to the original charge-conjugated trace in Eq.~\eqref{eq:trace0}; at an exact rational light-front test point, the original, standard, and covariant traces agree identically. The reduced light-front identity in Eq.~\eqref{eq:tracereduced} and the heavy good-component relation in Eq.~\eqref{eq:heavy} are also satisfied exactly.

The transverse integration measure was tested by replacing the LFQM integrand with a factorized Gaussian, $\exp(-a k_u^2-b k_d^2)$. The analytic integral with the measures used in Eq.~\eqref{eq:rho} is $I_{\rm G}=1/(256\pi^4ab)$. Removing one overall azimuth and, independently, applying the radial map in Eq.~\eqref{eq:kmap} reproduce this result with relative differences $3.79\times10^{-17}$ and $1.76\times10^{-17}$, respectively. The triangle map in Eq.~\eqref{eq:trianglemap} also gives $\int dx_u\,dx_d\,1=1/2$ exactly.

For the central parameter set, the unnormalized integral is \(N_{\rm raw}=2.761\times10^{-2}\). This number is quoted only as an implementation-level cross-check and carries no physical precision beyond that of the model inputs. After normalization, both the unit-integral condition and the longitudinal momentum sum rule agree with unity to better than \(10^{-14}\). The numerical \(u\leftrightarrow d\) exchange test is satisfied at the \(10^{-15}\) level, while the largest relative change between the last two transverse quadrature orders is \(4.5\times10^{-5}\). These numerical effects are therefore well below the model sensitivities discussed in Sec.~\ref{sec:results}.

\section*{Acknowledgements}
This work is supported by the Scientific Research Project of Colleges and Universities in Henan Province (No. 27AQ140025) and the Scientific Research Project of the Anhui Provincial Department of Education (Grant No. 2025AHGXZK40010).

\end{document}